\documentclass[%
 reprint, 
superscriptaddress,
 amsmath,amssymb,
 aps,
pra, 
longbibliography, floatfix ]{revtex4-1}

\usepackage{natbib} 
\usepackage{amsmath} 
\usepackage{graphicx} 
\usepackage{longtable}
\usepackage{multirow}
\usepackage{amsmath}
\usepackage{epstopdf}
\usepackage{placeins}

\usepackage{xcolor}  

\usepackage[english]{babel}

\makeatletter
\def\bbl@set@language#1{%
  \edef\languagename{%
    \ifnum\escapechar=\expandafter`\string#1\@empty
    \else\string#1\@empty\fi}%
  \@ifundefined{babel@language@alias@\languagename}{}{%
    \edef\languagename{\@nameuse{babel@language@alias@\languagename}}%
  }%
  \select@language{\languagename}%
  \expandafter\ifx\csname date\languagename\endcsname\relax\else
    \if@filesw
      \protected@write\@auxout{}{\string\select@language{\languagename}}%
      \bbl@for\bbl@tempa\BabelContentsFiles{%
        \addtocontents{\bbl@tempa}{\xstring\select@language{\languagename}}}%
      \bbl@usehooks{write}{}%
    \fi
  \fi}
\newcommand{\DeclareLanguageAlias}[2]{%
  \global\@namedef{babel@language@alias@#1}{#2}%
} \makeatother

\DeclareLanguageAlias{en}{english}

\begin{document}

\title{Exploring the Structure of Misconceptions in the Force and Motion Conceptual Evaluation with Modified Module Analysis}
\author{James Wells}
\affiliation{%
    College of the Sequoias, Science Division, Visalia CA, 93277
}%
\author{Rachel Henderson}
\affiliation{%
    Michigan State University, Department of Physics and Astronomy,
    East Lansing MI, 48824
}%
\author{Adrienne Traxler}
\affiliation{%
    Wright State University, Department of Physics,
    Dayton OH, 45435
}%
\author{Paul Miller}
\affiliation{%
    West Virginia University, Department of Physics and Astronomy,
    Morgantown WV, 26506
}%
\author{John Stewart}
\email{jcstewart1@mail.wvu.edu}
\affiliation{%
    West Virginia University, Department of Physics and Astronomy,
    Morgantown WV, 26506
}%

\date{\today}

\begin{abstract}

Investigating student learning and understanding of conceptual
 physics is a primary research area within Physics
Education Research (PER). Multiple quantitative methods have been
employed to analyze commonly used mechanics conceptual
inventories: the Force Concept Inventory (FCI) and the Force and
Motion Conceptual Evaluation (FMCE). Recently, researchers have
applied network analytic techniques to explore the structure of
the incorrect responses to the FCI identifying communities of
incorrect responses which could be mapped on to common
misconceptions. In this study, the method used to analyze the FCI,
Modified Module Analysis (MMA), was applied to a large sample of
FMCE pretest and post-test responses ($N_{pre}=3956$,
$N_{post}=3719$). The communities of incorrect responses
identified were consistent with the item groups described in
previous works. As in the work with the FCI, the network was
simplified by only retaining nodes selected by a substantial
number of students. Retaining nodes selected by 20\% of the
students produced communities associated with only four
misconceptions. The incorrect response communities identified for
men and women were substantially different, as was the change in
these communities from pretest to post-test. The 20\% threshold
was far more restrictive than the 4\% threshold applied to the FCI
in the prior work which generated similar structures. Retaining
nodes selected by 5\% or 10\% of students generated a large number
of complex communities. The communities identified at the 10\%
threshold were generally associated with common misconceptions
producing a far richer set of incorrect communities than the FCI;
this may indicate that the FMCE is a superior instrument for
characterizing the breadth of student misconceptions about
Newtonian mechanics.

\end{abstract}

\maketitle

\section{Introduction}

Understanding common difficulties students exhibit in learning
conceptual physics has been an important research strand in
physics education research (PER) since its inception. This work
was greatly advanced by the introduction of multiple-choice
conceptual instruments measuring students' understanding of
mechanics and electricity and magnetism: the Force Concept
Inventory (FCI) \cite{hestenes1992}, the Force and Motion
Conceptual Evaluation (FMCE) \cite{thornton1998}, the Conceptual
Survey of Electricity and Magnetism (CSEM) \cite{maloney2001}, and
the Brief Electricity and Magnetism Assessment (BEMA)
\cite{ding2006evaluating}. Studies involving these instruments
continue to be of central importance in PER. For an overview of
the history of these instruments and their use in PER, see Docktor
and Mestre's extensive synthasis of the field
\cite{docktor2014synthesis}.

Recently, substantial efforts have been made to apply quantitative
techniques to further understand these instruments including
factor analysis \cite{scott2012exploratory, semak2017,
eaton2018confirmatory},  cluster analysis
\cite{fazio2018conceptual}, and item response theory
\cite{wang2010, scott2015,stewart2018,zabriskie2019}. In 2016,
Brewe, Bruun, and Bearden \cite{brewe2016} introduced a new class
of quantitative algorithms to analyze the incorrect answers,
network analytic methods \cite{newman, zweig}. Network analysis is
a broad, flexible, and extremely productive field of quantitative
analysis that has been used to analyze systems as diverse as the
functional networks in the brain \cite{devico} and passing
patterns of soccer teams \cite{pena}.

A network is formed of nodes which are connected by edges. Network
analysis seeks to identify structure within the network; one
important class of structure is subsets of the network which are
more interconnected within themselves than they are connected to
the rest of the network. These subsets are called ``modules'' or
``communities'' interchangeably. In anticipation of the ``igraph''
package \cite{igraph} in the ``R'' software system
\cite{R-software} becoming the primary tool used within PER for
network analysis, we will call the subgroups ``communities.''

Wells {\it et al.} \cite{wells2019} attempted to replicate Brewe's
{\it et al.} \cite{brewe2016} analysis for the FCI and found that
it did not scale to large datasets. They suggested a modified
algorithm called Modified Module Analysis (MMA); the details are
discussed below as Study 1. In the current study, the MMA
algorithm was applied to explore the community structure of the
FMCE; the results are then compared to the results of
Study 1.

This study sought to answer the following research questions:
\begin{description}
\item[RQ1] What incorrect answer communities are identified by
Modified Module Analysis in the FMCE?  \item[RQ2] How are these
communities different pre- and post-instruction? How is the
community structure different for men and women? \item[RQ3] How do
the communities change as the parameters of the MMA algorithm are
modified?  \item[RQ4] How do the communities detected compare to
those detected in the FCI in Study 1?
\end{description}

\subsection{The FMCE Instrument}
\label{sec:fmce} The FMCE is a widely used mechanics conceptual
inventory that measures students' understanding of force and
motion. The instrument consists of 43 items examining student
understanding of Newton's laws of motion.  The items are presented
in groups with each item having at least 6 possible responses,
some of which represent common misconceptions. Most items include
a ``none of the above'' response which is not the correct response
to any item; ``none of the above'' responses have been shown to
cause psychometric problems \cite{devore_examining_2016}. The FMCE
is available at PhysPort \cite{physport}.

The FMCE uses the practice of ``blocking'' or ``chaining'' items
where multiple items refer to a common stem. In an item block, a
physical system is introduced, then multiple items refer to that
system. Of the 43 items in the FMCE, all but one (item 39) are
included in item blocks. The FCI also employs item blocks with 13
of the 30 items included in blocks. Multiple studies have
suggested that blocking items introduces spurious correlations
that can make the instrument difficult to interpret statistically
\cite{stewart2018,wells2019}.

Since its introduction, the blocked structure of the FMCE has been
used to provide a compact description of the instrument in terms
of the qualitative features of the item blocks. This description
has been refined since the introduction of the instrument as will
be discussed in Sec. \ref{sec:gen}. The descriptive terms provide
an overview of the instrument. ``Force Sled'' items (items 1-7)
ask about the force that an individual would need to exert on a
sled on a low-friction surface to produce a set of accelerations;
students select for a number of textual responses. ``Cart on a
Ramp'' items (items 8-10) ask students to select the force on a
cart as it moves up and down an incline. ``Coin Toss - Force''
items (items 11-13) ask students to select the force on a coin
tossed in the air. ``Force Graph'' items (items 14-21) ask
students about the force on a toy car as it moves across a
low-friction surface; students select from a number of graphs.
``Acceleration Graph'' items (items 22-26) ask students to select
the graph which correctly represents the acceleration of a toy car
moving on a horizontal surface. ``Coin Toss - Acceleration'' items
(items 27-29) ask students to select the acceleration of a coin
tossed in the air. ``Newton III'' items (items 30-39) ask students
about the forces during a variety of interactions between cars and
trucks. ``Velocity Graph" items (items 40-43) ask students to
select the graph which correctly represents the velocity of a toy
car moving on a horizontal surface. The current version of the
FMCE has four multiple choice ``Energy'' items (items 44-47) and
one free response item (46a). These items were not present in the
original FMCE and will not be analyzed in this study.

\subsection{Prior Studies}

As this analysis was motivated by prior works, this research will
draw heavily from two previous studies which will be referenced as
Study 1 and Study 2 throughout the manuscript.

\subsubsection{Study 1: Modified Module Analysis \label{Study1}} In
Study 1, Wells {\it et al.} \cite{wells2019} introduced Modified
Module Analysis (MMA), a network analytic method to explore the
structure of the incorrect answers of a multiple-choice
instrument. Modified Module Analysis was introduced to the adapt
Module Analysis of Multiple-Choice Responses (MAMCR) method of
Brewe {\it et al.} \cite{brewe2016} for a large datasets. In both
MMA and MAMCR, the incorrect responses to a conceptual inventory
are used to define a network with weighted edges. The responses
are the nodes of the network. In MAMCR, the number of times two
responses are selected by the same student define the edge weight
of the network. For example, if FCI response 1D and 2B were
selected together by 40 students, the network would contain 1D and
2B as nodes and have an edge between the nodes with weight 40. The
notation 1D represents response ``D'' to item 1. In MMA, the edge
weight is the correlation coefficient between the two responses.

To analyze this network, the correlation matrix was calculated and
a threshold applied. In Study 1, only edges which were correlated
at the $r>0.2$ level were retained where $r$ is the correlation
coefficient. The remaining correlated items define a network with
edge weight equal to the correlation. A community detection
algorithm was then applied to detect substructure in the network.
A community represents a set of nodes that are preferentially
selected together by many students. The MMA algorithm detects
incorrect answer communities, subsets of the network formed of
incorrect answers which are preferentially selected together.
Modified Module Analysis identified 9 pretest communities and 11
post-test communities on the FCI. Three of the communities were
the result of blocked items. For these blocked items, the later
response was the correct response if an earlier response had been
correct. In most cases, the remaining communities could be related
to the misconceptions associated with the items in original paper
introducing the FCI \cite{hestenes1992} and in the more detailed
taxonomy provided by Hestenes and Jackson \cite{fcitable}. For
eight of the communities, a dominant misconception was identified
and for two of the communities, two common misconceptions were
identified. For example, one FCI community included responses
\{4A, 15C, 28D\}, common incorrect answers to the Newton's 3rd law
items. Students were applying both the greater mass implies
greater force and the most active agent produces greater force
misconceptions for these items.

Study 1 found the communities identified for men and women on both
the pretest and post-test, while not identical, were very similar.

\subsubsection{Study 2: Multidimensional Item Response Theory and
the FMCE}

Study 1 made extensive use of a prior study of the FCI applying
constrained Multidimensional Item Response Theory (MIRT) to
produce a detailed model of the physical reasoning required to
correctly solve the items in the instrument \cite{stewart2018}.
The incorrect communities not related to the blocking of items
often required similar physical reasoning for their solution. This
methodology has recently been extended to the FMCE and will be
referenced as Study 2.  In Study 2, Yang {\it et al.} performed a
detailed analysis of the correct answers to the FMCE using
constrained MIRT \cite{fmce-mirt}. This technique produced a
detailed model of the instrument in terms of the fundamental
reasoning steps (principles) required for its solution. Results of
factor analysis and correlation analysis were also presented. All
analyses suggested the existence of subsets of items within the
instrument that shared a common solution structure. These item
groups included items 40-43 (definition of velocity), 22-26
(definition of acceleration), 30-39 (Newton's 3rd law), and 8-13
and 27-29 (motion under gravity). A fifth group of items, items
1-7 and 14-20, measured a combination of Newton's 1st and 2nd law
and corollaries of motion derived from these laws. These item
groups presented responses to students using different
representations with items 1-7 asking students to select textual
responses and items 14-20 asking students to choose between
two-dimensional graphs. The constrained MIRT analysis found that
this distinction between textual and graphical responses was
important to understanding student answers to the instrument.

The groups identified as requiring a common solution structure are
well aligned with the item groups identified by previous research
and described in Sec. \ref{sec:fmce} supporting the identification
of these groups as measuring distinct elements of Newtonian
thinking. Some of the groups suggested by MIRT combine groups
suggested by previous authors. For example, ``Cart on a Ramp,''
``Coin Toss - Force,'' and ``Coin Toss - Acceleration'' items all
require an understanding of the force or acceleration due to
gravity for their solution. Item groups with similar correct
solution structure will often also have responses that represent
consistently applied misconceptions in the analysis which follows.

In general, the FMCE had many more items requiring similar
reasoning for their solution than the FCI; this may make it a
productive instrument for the exploration of structure of
misconceptions about mechanics using MMA.

\section{Previous Studies of the FMCE}

\subsection{General Analyses}
\label{sec:gen}

Multiple subdivisions of the FMCE have been suggested. Thornton
and Sokoloff introduced four subgroups of items with the original
publication of the instrument: ``Force Sled'' items, ``Cart on a
Ramp'' items, ``Coin Toss'' items , and ``Force Graph'' items
\cite{thornton1998} as described above. Items 5, 6, and 15 were
identified as potentially problematic leading to modified
subgroups: ``Force Sled'' items (items 1-4 and 7) and ``Force
Graph'' items (items 14 and 16-21).

Using data collected after the instrument's publication, Thornton
\textit{et al.} proposed an alternate scoring scheme which
eliminated some items and scored some groups of items (clusters)
together \cite{thornton2009comparing}. The alternate scoring
scheme for the clusters suggested item groups 8-10, 11-13, and
27-29 be scored together because students had not mastered the
concept tested by the group unless they answered each item in the
group correctly. Each cluster received two points if all items
were answered correctly, zero points if not. They also suggested
the elimination of items 5, 15, 33, 35, 37, and 39 because
students without an understanding of Newtonian mechanics often
answered them correctly. They also suggested the elimination of
item 6 because content experts often answered it incorrectly.

Multiple authors proposed other revisions to the subgroups of
items initially introduced by Thornton and Sokoloff. Wittmann
identified five subgroups: ``Force (Newton I and II)'' (items 1-4,
7-14, 16-21), ``Acceleration'' (items 22-29), ``Newton III''
(items 30-32, 34, 36, 38), ``Velocity'' (items 40-43), and
``Energy'' (items 44-47) \cite{smith2008}. These subgroups were
further refined using a resource framework by Smith and Wittmann
who proposed a set of seven subgroups: ``Force Sled'' (items 1-4,
7), ``Reversing Direction'' (items 8-13, 27-29), ``Force Graphs''
(items 14, 16-21), ``Acceleration Graphs'' (items 22-26), ``Newton
III'' (items 30-32, 34, 36, 38), ``Velocity Graphs'' (items
40-43), and ``Energy'' (items 44-47) \cite{smith2008}. The
problematic items identified by Thornton \textit{et al.} were
eliminated from all subgroups in these two studies. More recently,
Smith, Wittmann, and  Carter applied the revised subgroup
structure to understand of the effect of instruction
\cite{smith2014}.

\subsection{Exploratory Analyses}
Many studies have applied quantitative analysis methods to explore
the structure of conceptual physics instruments. A substantial
number of studies have explored the factor structure of the FCI,
generally finding inconsistent or unintelligible results
\cite{huffman_what_1995,scott2012exploratory,scott2015,semak2017,stewart2018}.

Only two studies have performed factor analysis on the FMCE. Ramlo
examined the reliability of the FMCE using a sample of 146
students \cite{ramlo2008validity} finding adequate reliability on
the pretest (Cronbach's $\alpha=0.742$) and excellent reliability
on the post-test (Cronbach's $\alpha=0.907$). While the pretest
factor structure was undefined, three conceptually coherent
factors were identified on the post-test.

In Study 2, exploratory factor analysis found 5, 6, 9, and 10
factor models optimized some fit statistics. Overall, the model
fit of the 5-factor model was superior. The factor loadings in
this model were very consistent with the groups of conceptually
similar items identified by the confirmatory MIRT analysis.  These
groups also had adequate to excellent internally consistency
measured by Cronbach's alpha ranging from $\alpha=0.66$ to
$\alpha= 0.93$. There is also strong theoretical support for the
selection of either a 5 or 10 factor model as discussed in Study
2. Study 2 concluded that the 3-factor structure identified by
Ramlo probably resulted from the low sample size.

Recent studies of the FMCE have ranked incorrect responses to
examine conceptual development in introductory mechanics
\cite{smith2018showing} and produced a hierarchy of responses
\cite{Louis2019}.

\subsection{Gender and the FMCE}

On mechanics conceptual inventories (the FCI and the FMCE), men,
on average, outperform women by 13\% on the pretest and 12\% on
the post-test \cite{madsen2013}. The majority of research into the
``gender gap'' in PER analyzes differences between men and women
on the FCI; however, some studies have explored these differences
on the FMCE.

Researchers have explored various factors that could explain the
differences between men and women on the FMCE. For example,
differences in academic backgrounds and preparation, measured by
FMCE pretest and math placement exam scores, have been shown to
explain much of the gender gap on the FMCE post-test
\cite{kost2009,salehi2019}. Studies have also investigated the
impact of interactive-engagement on the overall gender gap.
Although some studies have shown a positive impact by reducing the
differences between men and women on conceptual inventory scores
\cite{lorenzo2006,kost2009,kohl2009introductory}, other
researchers have demonstrated that the gender gap for students
enrolled in an interactive-engagement classroom is unchanged
\cite{pollock2007}.

While many studies have focused on the overall average gender
differences on the FMCE, recently, researchers have explored the
fairness in the individual items on the FMCE \cite{henderson2018}.
An item is fair if men and women of equal overall ability with the
material score equally on the item. Applying the modified scoring
method proposed by Thornton \textit{et al.}
\cite{thornton2009comparing}, only item cluster 27-29 scored as a
single item consistently showed substantial unfairness in multiple
samples; this item was unfair to men. In one of the two samples,
item 40 demonstrated substantial gender unfairness; this item was
also unfair to men. These results were substantially different
from the analysis performed by Traxler \textit{et al.} which
identified a large number of unfair items on the FCI; most of the
items items were unfair to women \cite{traxler2018}.

\subsection{The FCI and the FMCE}

While both the FCI and the FMCE measure an understanding of
Newtonian mechanics, the FCI includes a substantially broader
coverage of the topic. The FCI includes two-dimensional kinematics
and circular motion while the FMCE does not. Thornton \textit{et
al.} \cite{thornton2009comparing} quantified this difference in
coverage noting that 22 of the 30 FCI items were outside the
coverage of the FMCE.

The optimal model presented in Study 2 and a similar study of the
FCI \cite{stewart2018} provide further evidence for the difference
in coverage of the two instruments with the optimal model of the
FCI requiring 19 principles (fundamental reasoning steps) while
the optimal model of the FMCE required only 8 principles. The two
instruments also differed starkly in their re-use of principles
with the FCI rarely repeating the same set of principles on
multiple items and the FMCE often repeating the same principles.
Study 2 also provided partial support for Thornton \textit{et al.}
\cite{thornton2009comparing} identification of problematic items
with items 5, 6, 33, 35, and 37 having relatively small
discriminations and item 15 having negative discrimination. The
models in Study 2 also suggest items 20 and 21 may not be
appropriately grouped with the other items probing graphical
interpretation of forces.

\section{The Structure of Knowledge}

The MMA algorithm detects sets of incorrect answers that are
commonly selected together by multiple students. Study 1 showed
that, for the FCI, these incorrect answer communities were related
to either misconceptions proposed by the authors of the FCI or to
the practice of blocking items. The reason students answer physics
questions incorrectly is a broad area of research and multiple
frameworks have been developed to explain incorrect answering.

\subsection{Knowledge Frameworks}
Much of the early work in PER conceptualized patterns of incorrect
answers as ``misconceptions,'' coherently applied incorrect
reasoning often related to Aristotelian or medieval theories of
nature. Early research investigated common student difficulties in
applying Newtonian mechanics
\cite{viennot1979,trowbridge1981,caramazza1981,peters1982,mccloskey1983,gunstone1987,camp1994}.
As the field evolved, systematic studies were developed to explore
student understanding and epistemology
\cite{mcdermott1997,thornton1998,rosenblatt2011,erceg2014,waldrip2014}.

Eventually, alternate frameworks not involving misconceptions were
proposed. Two of the most prominent frameworks are
knowledge-in-pieces \cite{disessa1993,disessa1998} and ontological
categories \cite{chi1993,chi1994,slotta1995}. Knowledge-in-pieces
models student thinking as resulting from the application of a set
of granular pieces of reasoning which are used independently or
collectively to solve problems. Multiple authors have investigated
this model and these reasoning pieces have been called
phenomenological primitives (p-prims)
\cite{disessa1993,disessa1998}, resources
\cite{hammer1996misconceptions,hammer_more_1996,hammer2000student},
and facets of knowledge \cite{minstrell1992}. In the
knowledge-in-pieces model, misconceptions represent consistently
activated p-prims. Unlike the misconception view, the
knowledge-in-pieces model views p-prims as potentially positive
resources than can be activated as part of the process of
constructing knowledge.

For a careful and accessible exploration of the relation of and
differences between the misconception view and the
knowledge-in-pieces framework, see Scherr \cite{scherr2007}; the
current study applies the definitions from this work. The
misconception model is defined as ``a model of student thinking in
which student ideas are imagined to be determinant, coherent,
context-independent, stable, and rigid'' \cite{scherr2007}. The
knowledge-in-pieces framework models student ideas ``as being at
least potentially truth-indeterminate, independent of one another,
context-dependent, fluctuating, and pliable'' \cite{scherr2007}.

The ontological category framework differs substantially from
either the misconception view or the knowledge-in-pieces view. The
ontological category framework models incorrect reasoning as
resulting for an incorrect classification of a concept. For
example, misclassifying force as a quantity that can be used up
which might lead a student to believe an object would slow when
the applied force was removed.

\subsection{Misconceptions}

The FCI was developed using the misconceptions model;  Hestenes,
Wells and Swackhamer proposed a detailed taxonomy of the
misconceptions measured by the instrument \cite{hestenes1992}. The
taxonomy was developed from qualitative studies investigating
students' ``alternate view of the relationship between force and
acceleration'' where researchers interviewed students about their
difficulties while solving conceptual physics problems
\cite{clement1982,clement1989,clement1993}. The authors of the FCI
provided a detailed description of the misconceptions measured by
the instrument \cite{hestenes1992}; this taxonomy was later
refined by Hestenes and Jackson \cite{fcitable}. The analysis in
the current work demonstrates that the FMCE probes a limited
number of the misconceptions that were originally outlined by the
authors of the FCI; only these misconceptions are described below.
For more information about the other misconceptions probed by the
FCI, see Study 1.

\vspace{6pt}

\noindent{\textit{Velocity-Acceleration Undiscriminated.}}  The
misconception of velocity-acceleration undiscriminated stems from
the concept of ``motion is vague'' \cite{hestenes1992}. This
misconception demonstrates the inability to differentiate the
concepts of position, velocity, and acceleration within
kinematics. For example, items 22-26 on the FMCE refer to a car
moving on a horizontal surface and ask for the acceleration as a
function of time. The velocity-acceleration undiscriminated
misconception would predict that when the car is speeding up or
slowing down at a constant rate, the graph would show a linear
trend of acceleration with respect to time and when the car is
traveling at a constant velocity, the graph would show a non-zero
constant acceleration.

\vspace{6pt}

\noindent{\textit{Motion Implies Active Forces.}} The motion
implies active forces misconception is one of the sub-categories
outlined under the ``Active Forces'' category of misconceptions
describe by the authors of the FCI \cite{hestenes1992}. This
misconception implies that an object in motion, even if moving at
constant velocity, will experience a force in the direction of
motion; it demonstrates that Newton's 2nd law is not well
understood. For example, items 1-4 on the FMCE probe this
misconception; a sled is being pushed along the ice and students
are asked to describe the force which would keep the sled moving.
The motion implies active forces misconception would predict that
force is proportional to velocity rather than acceleration.

\vspace{6pt}

\noindent{\textit{Action/Reaction Pairs.}} The misconceptions of
greater mass implies greater force and the most active agent
produces the greatest force are the two sub-categories within the
``Action/Reaction Pairs'' group of student difficulties. This
group of misconceptions implies that Newton's 3rd law is not well
understood. For example, FMCE items 30-32 probe these
misconceptions by describing collisions between a heavy truck and
a small car. The greater mass implies a greater force
misconception would predict that the heavy truck would exert a
greater force on the small car than the small car would on the
heavy truck. The most active agent produces the greatest force
would predict that the object that is moving the fastest would
produce the greatest force.

\section{Methods}

\subsection{Sample}

\label{sec:samples}

The sample was collected at a large eastern  land-grant university
serving approximately 30,000 students. The demographics of the
undergraduate population at the university were 80\% White, 6\%
International, 4\% African-American, 4\% Hispanic, 2\% Asian, 4\%
two or more races, and other groups less than 1\% \cite{usnews}.
The general undergraduate population had a range of ACT scores
from 21-26 (25th to 75th percentile).

The data were collected in the introductory calculus-based
mechanics course from Spring 2011 to Spring 2017. The majority of
the students enrolled in this course were physical science and
engineering majors. This sample was previously analyzed in
Henderson \textit{et al.} (Sample 3A \cite{henderson2018}) where
the instructional environment is described in detail. The course
was taught by multiple instructors and generally featured an
interactive pedagogy in lecture and laboratory.

Over the period studied, the FMCE was given at the beginning and
at the end of the class in each semester. The sample contains 3956
FMCE pretest responses and 3719 FMCE post-test responses (each
with 80\% men); only the students who completed the course for a
grade were included in the study. The overall pretest to post-test
gains for men and women were 28\% and 21\%, respectively. The
descriptive statistics for the FMCE pretest and the FMCE post-test
are presented in Table II in Henderson \textit{et al.} (Sample 3A)
\cite{henderson2018}.

\subsection{Analysis Methods}\label{sec:ModuleAn}


This work applies Modified Module Analysis (MMA) described in
Study 1 to the FMCE. Although the method is described in detail in
Study 1 \cite{wells2019}, we provide an overview of the method
here.

All responses to the FMCE where dichotomously coded where response
1D$_i$ would be coded as one if student $i$ selected the response
and zero otherwise. The correct responses were eliminated; network
analysis is unproductive if the correct responses are included
because they form a single tightly connected community that hides
the structure of the incorrect answers. Responses that were
selected by fewer than 5\% of the students were eliminated as
statistically unreliable.

The correlation matrix was calculated for the remaining incorrect
answers. This correlation matrix defines a network with nodes
representing the incorrect responses and weighted edges between
the nodes representing the strength of the correlation between the
two responses. Edges that represent correlations that were not
significant at the $\alpha = 0.05$ level with a Bonferroni
correction applied were eliminated. The network was further
simplified by eliminating any correlation where $r<0.2$; this was
the threshold applied in Study 1. This also served to remove the
large negative correlations between two responses to the same
item. Network analysis often uses methods to simplify the network
while retaining important structure; this process is called
``sparsification.''

A community detection algorithm was then applied to detect
structure in the network. Study 1 applied the ``fast-greedy''
algorithm \cite{newmangirvin:2004} included in the ``igraph''
package \cite{igraph} for R. Many community detection algorithms
exist; Study 1 reported that most produced similar results for the
correlation network. The fast-greedy algorithm is designed to
maximize the modularity of the division of the network into
unified subnetworks. Modularity measures the number of
intra-community edges in a particular division of the network
compared to the number expected in a random division.

To account for randomness in both the sample and the algorithm,
1000 bootstrapped replications were carried out. As a result, 1000
divisions of the network into communities were calculated sampling
the data with replacement. For each pair of incorrect items, the
number of times the two items appeared in the same community was
calculated. This number is divided by the number of bootstrap
replications to form the community fraction $C$. In this study, we
analyzed communities that were identified in 80\% of the 1000
bootstrapped samples.

Because the incorrect answer communities of men and women are
compared and the number of men in the sample is significantly
larger than the number of women, care was taken to produce a
balanced sample. For men, the data were downsampled to the size of
the female dataset. For women, the dataset was sampled with
replacement preserving the size of the dataset.

\section{Results}

Modified Module Analysis was applied to the FMCE; the communities
identified are shown in the first table in the Supplemental
Materials \cite{supp}. Retaining nodes where at least 5\% of the
students selected the response (approximately the threshold used
in Study 1) produced  35 communities. These communities were often
formed of small subsets of item groups identified in previous
studies. This was dramatically different than the small number of
communities identified in the FCI by Study 1. The complex nature
of the communities identified made understanding their structure
difficult.

To produce a simpler structure more open to interpretation, the
network was further sparsified retaining only nodes selected by
20\% of the students. The community structure of this network is
shown in Table \ref{tab:commat}. In nearly every case, the
communities form completely disconnected, complete graphs. The
intra-community density measures the connectivity of a community
and is defined as $\gamma = 2m/n(n-1)$, where $n$ is the number of
nodes and $m$ is the number of realized edges. A fully connected
community has an intra-community density of one.

\begin{table*}[!htb]
        \caption{Communities identified in the pretest and post-test incorrect answers at $r>0.2$ and community fraction, $C>0.8$. Only nodes selected by $20\%$ of the students are included. The number in parenthesis is the
        intra-community density, $\gamma$,
        for communities where the intra-community density is not one. Newton III* denotes that this community does not contain 31F. \label{tab:commat} }
    \centering
    \begin{tabular}{|l|cc|cc|c|}
        \hline
\multirow{2}{*}{Community}&\multicolumn{2}{|c}{Pretest}&\multicolumn{2}{|c|}{Post-test}&Item\\
&Men&Women&Men&Women&Group\\\hline
1A, 2B, 3C, 4G, 5B, 6C, 7E&X&X&&X&Force Sled\\\hline
\multirow{2}{*}{1A, 2B, 3C, 4G, 5B, 6C, 7E, 14A, 16C, 17B, 18H, 19D, 20F}&&&\multirow{2}{*}{X($\gamma = 0.88$)}&&Force Sled\\
&&&&&Force Graph\\\hline
\multirow{2}{*}{8G, 9D, 10B, 11G, 12D, 13B}&\multirow{2}{*}{X}&&\multirow{2}{*}{X}&&Cart on a Ramp\\
&&&&&Coin Toss - Force\\\hline
\multirow{3}{*}{8G, 9D, 10B, 11G, 12D, 13B, 27G, 28D, 29B} &&\multirow{3}{*}{X}&&\multirow{3}{*}{X}&Cart on a Ramp\\
&&&&&Coin Toss - Force\\
&&&&&Coin Toss - Acceleration\\\hline
14A, 16C, 17B, 18H, 19D, 20F, 21H &X&&&&Force Graph\\\hline
14A, 16C, 17B, 18H, 19D&&X&&X&Force Graph\\\hline
\multirow{2}{*}{22E, 23G, 24B, 25F, 26A, 27G, 28D, 29B}&\multirow{2}{*}{X}&&&&Acceleration Graphs\\
&&&&&Coin Toss - Acceleration\\\hline
22E, 23G, 24B, 25F, 26A&&X&X&X&Acceleration Graphs\\\hline
27G, 28D, 29B&&&X&&Coin Toss - Acceleration\\\hline
30A, 31F, 32B, 34B, 36C, 38B, 39D&X&&&&Newton III\\\hline
30A, 31F, 32B, 34B, 36C, 38B&&X&&&Newton III\\\hline
30A, 32B, 34B, 35B, 36C, 38B, 39D&&&X&X&Newton III*
\\\hline
    \end{tabular}
\end{table*}

\begin{figure*}[!htb]
    \centering
    \includegraphics[width=\textwidth]{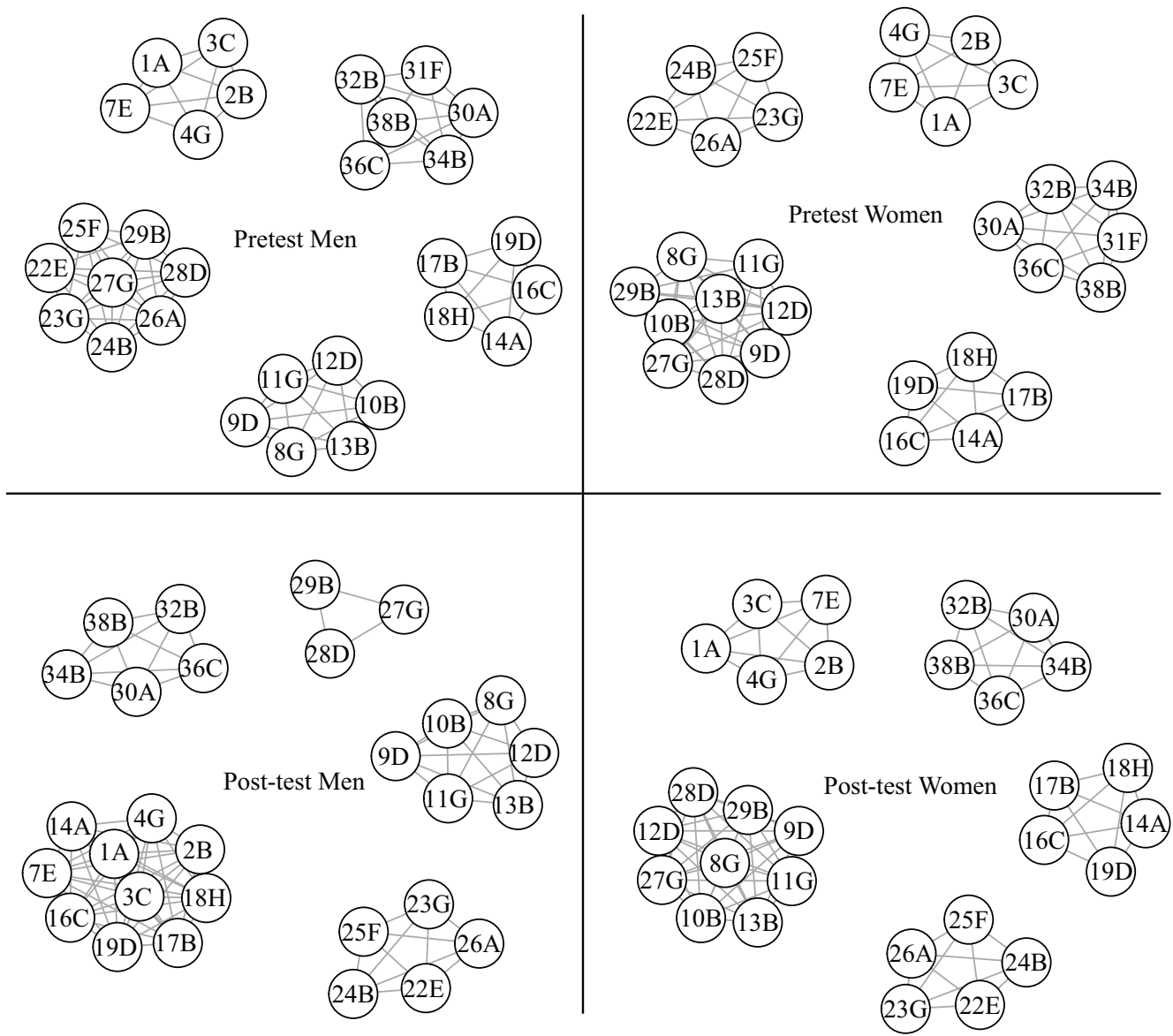}
    \caption{Communities identified in the FMCE pretest and post-test for men and women.\label{fig:network}}
\end{figure*}

Table \ref{tab:commat} offers partial support for the
identification of items 5, 6, 15, 33, 35, 37, and 39 as
problematic in Thornton \textit{et al.}
\cite{thornton2009comparing}. Items 20 and 21 were modeled as
having a different solution structure to other items in the
``Force Graph'' group in Study 2; these items are inconsistently
connected to the other items in this group in Table
\ref{tab:commat}. Incorrect answers to items 15, 33, and 37 were
never identified as part of a community. Incorrect answers to
items 20, 21, 35, and 39 were inconsistently identified as parts
of the communities associated with the items in the group. As
such, some of the complexity in Table \ref{tab:commat} results
from these items. If items 5, 6, 15, 20, 21, 33, 35, 37, and 39
are eliminated from the analysis, the structure of Table
\ref{tab:commat} simplifies substantially to produce Table
\ref{tab:commat2}. The communities in Table \ref{tab:commat2} are
shown graphically in Fig. \ref{fig:network}.

\begin{table*}[!htb]
        \caption{Communities identified in the pretest and post-test incorrect answers at $r>0.2$ and community fraction, $C>0.8$. Only nodes selected by $20\%$ of the students are included. Problematic items identified
        in Study 1 and 2 have been eliminated. The number in parenthesis is the
        intra-community density, $\gamma$,
        for communities where the intra-community density is not one. \label{tab:commat2} }
    \centering
    \begin{tabular}{|l|cc|cc|c|}
        \hline
\multirow{2}{*}{Community}&\multicolumn{2}{|c}{Pretest}&\multicolumn{2}{|c|}{Post-test}&Item\\
&Men&Women&Men&Women&Group\\\hline
1A, 2B, 3C, 4G, 7E&X&X&&X&Force Sled\\\hline
\multirow{2}{*}{1A, 2B, 3C, 4G, 7E, 14A, 16C, 17B, 18H, 19D}&&&\multirow{2}{*}{X($\gamma = 0.88$)}&&Force Sled\\
&&&&&Force Graph\\\hline
\multirow{2}{*}{8G, 9D, 10B, 11G, 12D, 13B}&\multirow{2}{*}{X}&&\multirow{2}{*}{X}&&Cart on a Ramp\\
&&&&&Coin Toss - Force\\\hline
\multirow{3}{*}{8G, 9D, 10B, 11G, 12D, 13B, 27G, 28D, 29B} &&\multirow{3}{*}{X}&&\multirow{3}{*}{X}&Cart on a Ramp\\
&&&&&Coin Toss - Force\\
&&&&&Coin Toss - Acceleration\\\hline
14A, 16C, 17B, 18H, 19D&X&X&&X&Force Graph\\\hline
\multirow{2}{*}{22E, 23G, 24B, 25F, 26A, 27G, 28D, 29B}&\multirow{2}{*}{X}&&&&Acceleration Graphs\\
&&&&&Coin Toss - Acceleration\\\hline
22E, 23G, 24B, 25F, 26A&&X&X&X&Acceleration Graphs\\\hline
27G, 28D, 29B&&&X&&Coin Toss - Acceleration\\\hline
30A, 31F, 32B, 34B, 36C, 38B&X&X&X&X&Newton III\\\hline
    \end{tabular}
\end{table*}

The sets of items in Table \ref{tab:commat} and \ref{tab:commat2}
generally conform to the item groups identified in previous works
and discussed in Sec. \ref{sec:fmce}. Table \ref{tab:commat2}
suggests items 27-29 should be treated as an independent group; we
propose this group be called ``Coin Toss - Acceleration'' to
distinguish it from items 11-13 which becomes ``Coin Toss -
Force.'' Both sets of items ask about a coin tossed in the air;
items 11-13 ask about the force on the coin, items 27-29 about the
acceleration. Smith and Wittmann combined these items into a
``Reversing Direction'' (items 8-13, 27-29) group; MMA suggests
this grouping may not be appropriate for all students. We also
note that Smith and Wittmann's ``Velocity Graphs'' (items 40-43)
group does not appear. This group had relatively poor Cronbach
alpha when used as a subscale in Study 2.

At this level of sparsification, for each item only a single
response appeared in each community, indicating that there is a
single, dominant incorrect answer that students tend to select.
This was consistent between the pretest and the post-test and by
gender.

\begin{table*}[t!]
    \caption{Item groups, the physical principle tested by the group, and the common misconception selected by the students. \label{tab:summary} }
    \centering
    \begin{tabular}{|l|c|l|l|}
        \hline
        Item Group & Community & Physical Principle &
        Misconception\\\hline
        Force Sled&1A, 2B, 3C, 4G, 7E&Newton's 1st and 2nd law&Motion implies active forces\\\hline
        Cart on a Ramp&8G, 9D, 10B&Motion under gravity&Motion implies active forces\\\hline
        Coin Toss - Force&11G, 12D, 13B&Motion under gravity&Motion implies active forces\\\hline
        Force Graph&14A, 16C, 17B, 18H, 19D&Newton's 1st and 2nd law&Motion implies active forces\\\hline
        Acceleration Graphs&22E, 23G, 24B, 25F, 26A&Definition of acceleration&Velocity-acceleration undiscriminated\\\hline
        Coin Toss - Acceleration&27G, 28D, 29B&Motion under gravity&Velocity-acceleration undiscriminated\\\hline
        \multirow{2}{*}{Newton III}&\multirow{2}{*}{30A, 31F, 32B, 34B, 36C, 38B}&\multirow{2}{*}{Newton's 3rd law}&Greater mass implies greater force\\
        &&&Most active agent produces greatest force\\\hline
    \end{tabular}
\end{table*}

\subsection{The Structure of Incorrect FMCE Responses}

Study 2 allows the description of the physical principles tested
by each item group. Both ``Force Sled'' and ``Force Graph''  test
a combination of Newton's 1st and 2nd law and the definition of
acceleration. The ``Force Graph'' items also require the use of
graphical reasoning. The ``Cart on a Ramp,'' ``Coin Toss -
Force,'' and ``Coin Toss - Acceleration'' groups each require the
law of gravitation, that the gravitational force is downward and
constant. The ``Acceleration Graphs'' group requires the
definition of acceleration and reading a graph. The ``Newton III''
group requires Newton's 3rd law.

In addition to the communities being strongly related to the
item groups, often multiple item groups
testing the same physical principles were part of the same
community. Much of the complexity of Table \ref{tab:commat2}
results from the inconsistent joining of incorrect answers to
items testing the same concept. Table \ref{tab:summary} summarizes
the item groups, the physical principle tested by the group, and
the common misconception selected for the group.

The misconceptions represented by the items in the incorrect
communities are quite consistent. As in Study 1, we use Hestenes
and Jackson's extensive taxonomy of misconceptions measured by the
FCI to classify the misconceptions \cite{fcitable}. The ``Force
Sled,'' ``Force Graph,'' ``Coin Toss - Force,'' and ``Cart on a
Ramp'' responses all represent the motion implies active forces
misconception; all select a force proportional to the velocity.
The ``Acceleration Graphs'' and ``Coin Toss - Acceleration''
groups both represent the velocity-acceleration undiscriminated
misconception; all select an acceleration proportional to
velocity.

Study 1 found that the FCI presented the students with two
misconceptions related to Newton's 3rd law: greater mass implies
greater force and most active agent produces greatest force. MMA
was unable to disentangle the application of these two
misconceptions for the FCI. Both misconceptions are also in the
same community for the FMCE. Item 30A represents the greater mass
implies greater force misconception. Items 32B, 34B, 36C, 38B
apply the most active agent produces greatest force misconception.
Interestingly, item 31 gives the student a situation where both
misconceptions apply, a head-on collision between a large truck
and a faster moving car. Response 31F indicates the student does
not believe they have enough information to solve the item
suggesting they are indeed trying to apply both misconceptions
simultaneously.

\subsection{Gender Differences in Community Structure}

Both men and women consistently answer incorrectly to the ``Force
Sled'' and ``Force Graph'' items on the pretest. The physical principles
needed to solve these items are very similar, but the responses to
the ``Force Sled'' items are textual whereas the responses to the
``Force Graph'' items are graphical. This seems to indicate that the
representation chosen for the answer affects the application of
the misconception on the pretest for both men and women. These item
groups continue to be different communities for women on the
post-test; for men, they have generally merged ($\gamma=0.88$)
into a single community on the post-test.

Men and women also differ in their application of misconceptions
to items involving motion under gravity: ``Cart on a Ramp'' items,
``Coin Toss - Force'' items, and ``Coin Toss - Acceleration''
items. These items form a single community on both the pretest and
post-test for women. For men, the Coin Toss - Acceleration items
are in a different community on both the pretest and post-test.
These three groups do apply different misconceptions with ``Cart
on a Ramp'' and ``Coin Toss - Force'' items applying a force
proportional to velocity misconception while the ``Coin Toss -
Acceleration'' items apply an acceleration proportional to
velocity misconception. If a student understands that force and
acceleration are proportional, then these two misconceptions
should produce the same results. The pattern of community
membership seems to indicate women apply both misconceptions
consistently, while men do not.

While most communities make theoretical sense, both in terms of
the item group suggested for the instrument and the physical
principles required to solve items in the group identified in
Study 2, one does not. For men, one pretest community combines
``Acceleration Graphs'' with ``Coin Toss - Acceleration.'' These
items require very different physical reasoning for their correct
solution, but apply the same misconception, velocity-acceleration
undiscriminated. For these items, the misconception is more
important in determining the community than the correct answer
structure.

\subsection{The Strength of Common Misconceptions}

\begin{table*}[t]
        \caption{\label{tab:misc} Percentage of students selecting each incorrect community for the FMCE post-test; mean,
        1st quartile (1Q), median (med), and 3rd quartile (3Q). A Mann-Whitney $U$ test was performed to
        determine if the differences between men and women were significant, the $p$-value is presented.
        The effect size is given as Vargha and Delaney's $A$ \cite{vargha2000},
        the probability that a randomly selected woman will score higher than a randomly selected man.
        }
    \centering
    \begin{tabular}{|l|cc|cc|cc|c|}\hline
    \multirow{2}{*}{Community}&\multicolumn{2}{|c}{Men}&\multicolumn{2}{|c|}{Women}&\multirow{2}{*}{$p$}&\multirow{2}{*}{$A (\%)$}&\multirow{2}{*}{Misconception}\\
    &Mean&1Q, Med, 3Q &Mean&1Q, Med, 3Q (\%)&&&\\\hline
   Force Sled, Force Graph                                          & 48        &$10, 50,  80$  &59&$40, 70,  80$&$<0.001$&59&Motion implies active forces\\\hline
   Cart on a Ramp                               &\multirow{2}{*}{48}         & \multirow{2}{*}{$0, 50, 83$}   &\multirow{2}{*}{59}& \multirow{2}{*}{$33, 67,   83$}&\multirow{2}{*}{$<0.001$}&\multirow{2}{*}{61}&\multirow{2}{*}{Motion implies active forces}\\
   Coin Toss - Force &&&&&&&\\\hline
   Acceleration Graphs                                              & 27        &$0, 0,  60$    &35& $0, 20,  60$&$<0.001$&56&Velocity-acceleration undiscriminated\\\hline
   Coin Toss - Acceleration                                         & 30        &$0, 0,  67$    &44& $0, 33,  67$&$<0.001$&62&Velocity-acceleration undiscriminated\\\hline
   \multirow{2}{*}{Newton III} & \multirow{2}{*}{43}        &\multirow{2}{*}{$0, 40,  80$}   &\multirow{2}{*}{46}& \multirow{2}{*}{$0, 40,  80$}&\multirow{2}{*}{$0.07$}&\multirow{2}{*}{52}&Greater mass implies greater force\\
   &&&&&&&Most active agent produces largest force\\

    \hline
    \end{tabular}
\end{table*}

One potential application of these results is to provide classroom
instructors with a measurement of how strongly a misconception is
held by their students. The instructor could then tailor his or
her instruction to emphasize material on those subjects. The
strength of a misconception community, called the ``misconception
score,'' is defined as the fraction of items within the community
that are selected by the student. For example, if a community
contains \{22E, 23G, 24B, 25F, 26A\}, a student who selected 22E,
24G, and 26A would have a misconception score of sixty percent,
while a student who selected all five answer choices would have a
score of one-hundred percent. A higher score indicates a more
strongly held misconception. A student who answered items 22, 23,
24, 25, and 26 correctly would have a misconception score of zero
percent.

The Mann-Whitney $U$ test \cite{mann1947} was used to determine if
the misconception scores were significantly different for men and
women on the post-test because the data were highly non-normal and
discontinuous. The Mann-Whitney $U$ test is a non-parametric test
that may be used instead of the unpaired t-test. In this sample,
the overall post-test score was higher for men than women: the
median number of incorrect responses was 20 for men and 26 for
women. The effect size of this difference, measured using Vargha
and Delaney's $A$ statistic \cite{vargha2000}, was small: 0.63.
This indicates that a randomly selected female student will have
more incorrect answers than a randomly selected male student 63\%
of the time. If there were no effect, $A$ would be 0.50,
reflecting a 50-50 chance of a score from either group being
higher. The small, medium, and large effect sizes for Cohen's $d$
correspond to values of Vargha and Delaney's $A$ greater than
0.56, greater than 0.64, and greater than 0.71, respectively.

Table \ref{tab:misc} presents the $A$ statistic, the mean, 1st
quartile (1Q), median (Med.), and third quartile (3Q) for men and
women for the misconception scores for each incorrect answer
community. While the Mann-Whitney $U$ test found a significant
difference in each case, all of the $A$ values were in the small
or negligible effect size range. Furthermore, all of the $A$
values were lower than the overall chance of selecting a female
student at random with more incorrect answers than a random male
student. This is consistent with the finding in Study 1 showing
while significant differences exist between the misconception
scores of men and women, that these differences are largely
explained by overall differences in the post-test scores of men
and women.

For the class studied, students hold the motion implies active forces and the Newton's
3rd law misconceptions more strongly than the velocity-acceleration undiscriminated
misconception.

\subsection{Reducing Sparsification}

Sparsification is a network analytic term for removing edges from
a network to reduce its density. In MMA, sparsification is
accomplished by removing nodes selected by a small number of
students and edges correlated below some threshold ($r<0.2$ in
this study). Sparsification allows important structure to be
identified in the network. Table \ref{tab:commat2} presents the
community structure identified after sparsifying the network by
removing all nodes selected by fewer than $20\%$ of the students.
This sparsification results in a community structure very similar
to that identified in Study 1 with a small number of communities
each associated with a misconception discussed in Hestenes and
Jackson's \cite{fcitable} taxonomy.

This sparsification threshold is far more strict than that applied
in Study 1 which only removed nodes not selected by 30 students
(about 4\% of the sample). When a similar threshold was applied to
the FMCE, $5\%$, 35 communities were found in either the pretests
or post-tests of men and women. These results are presented in the
Supplemental Material \cite{supp}. Most of these communities were
very similar to one another, differing by only a single response
in some cases. These differences may have resulted from the very
different manner in which the two instruments treat incorrect
responses. The FCI presents the student with a number of responses
developed from student interviews, most designed to test a
specific misconception. Most students select only one or two of
the available incorrect answers. The FMCE presents the students
with many possible options that come close to exhausting the
available responses.

This greater scope of possible answers produces a more complex
community structure that offers the possibility of identifying
misconceptions not explicitly used to construct the instrument.
The communities identified for men and women on the pretest and
post-test for responses selected by a minimum of $10\%$ of the
students are also presented in the Supplemental Material
\cite{supp}. The misconceptions represented by communities not
identified at $20\%$ sparsification are shown in Table
\ref{tab:commat10misc}. While some responses do not have an
obvious relation to the general misconception tested by the
community (marked with an *), most responses in the communities
can be associated with a single misconception. Often these
misconceptions are outside the taxonomy \cite{fcitable} developed
for the FCI suggesting students have a much richer set of
misconceptions than is measured by the FCI. In Table
\ref{tab:commat10misc}, misconceptions identified by Hestenes and
Jackson \cite{fcitable} are bolded. Many of the items represent
combinations of misconceptions in this taxonomy involving the
failure to discriminate force, acceleration, velocity, and
position in varying combinations. The items mix the
position-velocity undiscriminated, the velocity-acceleration
undiscriminated, and the velocity proportional to applied force
misconceptions identified by Hestenes and Jackson \cite{fcitable}.
    \begin{table*}[t]
        \caption{Misconceptions represented by communities identified in items selected by at least 10\% of the students which were not identified in items selected
            by at least 20\% of the students. Items marked * do not have an obvious relation to the misconception. Misconceptions identified by Hestenes
and Jackson \cite{fcitable} are bolded.\label{tab:commat10misc} }
        \centering
        \begin{tabular}{|l|l|}
            \hline
            Community & Misconception\\\hline
            3D, 7D &No force is required to slow an object.\\\hline
            \multirow{2}{*}{3E, 7C}  &To slow an object at a constant rate, a decreasing force \\
            &opposite motion must be applied.\\\hline
            \multirow{2}{*}{3G, 7A}  &To slow an object at a constant rate, an increasing force \\
            &opposite motion must be applied.\\\hline
            8E, 11E, 27E&Gravity exerts a constant force in the direction of motion.\\\hline
            8F, 11F, 27F&Gravity exerts an increasing force in the direction of motion.\\\hline
            \multirow{2}{*}{8F, 10C, 11F, 13C, 27F, 29C} &Gravity exerts an increasing force as an object travels upward \\
            &and a decreasing force as it travels downward.\\\hline
            \multirow{2}{*}{8F, 10C, 11F}&Gravity exerts an increasing force as an object travels upward \\
            &and a decreasing force as it travels downward.\\\hline
            11E, 27E&Gravity exerts a constant force in the direction of motion.\\\hline
            14C, 17H, 24G, 26E, 40D, 42C, 43A*  &Force-acceleration-velocity undiscriminated from position.\\\hline
            14C, 17D, 17H, 23D*, 24G, 26E, 40D, 42C, 43A* &Force-acceleration-velocity undiscriminated from position.\\\hline
            14C, 17D, 40D, 42C &Force-velocity undiscriminated from position.\\\hline
            14C, 17D, 17H, 40D, 42C, 42H*&Force-velocity undiscriminated from position.\\\hline
            17A, 18D, 19C, 19H, 23F, 24A, 25E, 25G &{\bf Velocity proportional to applied force.}\\\hline
            17A, 19C, 24A, 25E, 42A* &{\bf Velocity proportional to applied force.}\\\hline
            18D, 19H, 23F, 25G&{\bf Velocity proportional to applied force.}\\\hline
            19C, 25E&{\bf Velocity proportional to applied force.}\\\hline
            24F, 26E&\textbf{Velocity-acceleration undiscriminated.}\\\hline
            \multirow{2}{*}{27B, 27C, 29F}&Gravitational acceleration not constant and in the opposite \\
            &direction of motion. \\\hline
            \multirow{2}{*}{27C, 29F}  &Gravitational acceleration proportional to velocity and in the opposite\\
            &direction of motion.\\\hline
        \end{tabular}
    \end{table*}

\section{Discussion}
\subsection{Research Questions}
This study sought to answer four research questions; the first
three will be addressed in the order proposed. The fourth research
question compares the results of Study 1 for the FCI to the
results of this study. The differences of the FCI and FMCE will be
discussed as part of the answer to each of the first three research
questions.

\textit{RQ1: What incorrect answer communities are identified by
Modified Module Analysis in the FMCE? } The communities of
incorrect responses identified on the FMCE generally conformed to
the  block structure of the instrument and were associated with
items groups identified in previous work. This discussion will
focus on the analysis retaining nodes selected by 20\% of the
students; results retaining nodes selected by 5\% and 10\% of the
students are discussed in RQ3. Modified Module Analysis showed the
item groups proposed by Smith and Wittman were being consistently
answered using a common misconception: the ``Force Sled'' (items
1-4, 7), the ``Force Graph'' (items 14, 16-19), ``Acceleration
Graphs'' (items 22-26) and ``Newton III'' (items 30-32, 34, 36,
38) \cite{smith2008}. The ``Reversing Direction'' subgroup of
items (items 8-10, 11-13, 27-29) \cite{smith2008} was not
consistently identified as an incorrect answer community. The
subgroup of items 27-29 sometimes formed its own community and was
sometimes grouped with the other items. We proposed renaming the
subgroups: ``Cart on a Ramp'' (items 8-10), ``Coin Toss-Force''
(items 11-13), and ``Coin Toss-Acceleration'' (items 27-29).
``Cart on a Ramp'' and ``Coin Toss - Force'' items were identified
in the same community both pre- and post-instruction and for men
and women; ``Coin Toss - Acceleration'' items were inconsistently
identified as part of this community.

Only four misconceptions were identified retaining nodes selected
by the 20\% of the students: motion implies active forces,
velocity-acceleration undiscriminated, and two Newton's 3rd law
misconceptions. The Newton's 3rd law misconceptions, greater mass
implies greater force and most active agent produces largest
force, were not identified as independent incorrect answer
communities. This is consistent with Study 1 which also failed to
distinguish the two misconceptions in the FCI. Also consistent
with Study 1, the incorrect answer communities contained items
testing the same physical principles as identified in Study 2. The
physical principle tested by the item, rather than the
misconception, was the most important factor in determining the
incorrect answer community. In this study, four separate item
groups were associated with the motion implies active forces
misconception (Table \ref{tab:summary}): ``Force Sled,'' ``Force
Graph,'' ``Cart on a Ramp,'' and ``Coin Toss - Force.'' Study 2
showed that the first two groups required Newton's 1st and 2nd law
for their solution while the last two required the law of
gravitation. While testing the same misconception, the first two
groups were never detected in the same community as the last two
groups. This is consistent with Study 1 which also identified
multiple incorrect answer communities in the FCI measuring the
motion implies active forces misconception; these communities also
had similar correct solution structure \cite{stewart2018}.

Study 2 demonstrated that the FMCE has substantially less complete
coverage of mechanics than the FCI which was consistent with
previous work by Thornton {\it et al.}
\cite{thornton2009comparing}. The FCI also measures a broader set
of misconceptions than the FMCE. Communities associated with 9
different misconceptions were identified in the FCI, while only 4
were identified in the FMCE. While covering fewer misconceptions,
the FMCE does measure the critical velocity-acceleration
undiscriminated misconception more thoroughly than the FCI.
Responses 19A, 20B, and 20C in the FCI are reported to measure
this misconception in Hestenes and Jackson \cite{fcitable}, but
were not detected as an incorrect answer community in Study 1.

Study 1 also identified 3 communities in the FCI that directly
resulted from the blocked structure of the instrument. In these
communities, the second item in an item block was the correct
answer if the first answer had been the correct answer. No such
communities were identified in the FMCE. While extensively
blocked, the items in the FMCE do not directly refer to the
results of previous items.

The communities identified in the FMCE were generally
substantially larger than those identified in the FCI. The FCI
contained 13 distinct communities for a 30-item instrument while
the FMCE contained 9 communities for a 43-item instrument. In the
FMCE, some of the distinct communities resulted from joining other
communities. All communities in the FMCE can be formed of 6 groups
of items: ``Force Sled,'' ``Force Graph,'' ``Acceleration
Graphs,'' ``Coin Toss - Acceleration,'' ``Newton III,'' and a
community that combines ``Cart on a Ramp'' and ``Coin Toss -
Force.'' As such, substantially fewer distinct groups of
misconceptions are identified in the FMCE; however, the groups
were often substantially larger in the FMCE than the FCI. For the
FMCE, the fundamental groups have sizes ranging from 3 to 6 with
all but one group containing at least 5 items. Only 2 of the 13
groups in the FCI contain as many as 3 items with 11 groups
containing only two items. Because the incorrect answer
communities contain more items, the FMCE may provide a
substantially more accurate characterization of the strength of
the misconception (Table \ref{tab:misc}) than the FCI.

The MMA method also provided support for eliminating the
problematic items which were identified by Thornton \textit{et
al.} \cite{thornton2009comparing}. With items 5, 6, 15, 20, 21,
33, 35, 37 and 39 included in the analysis, the community
structure was complex which made it rather difficult to interpret
because some of these items were inconsistently associated with a
misconception community.

\textit{RQ2: How are these communities different pre- and
post-instruction? How is the community structure different for men
and women?} The pre- and post-instruction differences of the
community structure were very different for men and women, and as
such, these two questions will be addressed together. The
communities identified for men and women were often different; on
the FMCE pretest, only three out of the nine communities were the
same, while on the FMCE post-test, two out of the nine were the
same. The differences were generally the result of joining two
communities with similar correct solution structure as identified
in Study 2. Men integrated the ``Force Sled'' and ``Force Graph''
item groups on the post-test while women did not; however, women
integrated the ``Coin Toss - Acceleration'' item group with the
``Cart on a Ramp'' and ``Coin Toss - Force'' item groups on the
post-test while men did not. As such, neither men nor women were
more likely to form more integrated misconceptions with
instruction. The same physical reasoning is required to solve the
items in the larger integrated misconception groups and,
therefore, more consistency in selecting a misconception may
represent progress in recognizing the same reasoning is required
by the items.

The difference between men and women both pre- and
post-instruction was dramatically different than the results of
Study 1 for the FCI. Generally, the incorrect answer community
structure was very similar for men and women on both the pretest
and the post-test for the FCI.

The change in misconception structure between the pretest and the
post-test was dramatically different for men and women. For women,
the misconception communities identified were completely
consistent from the pretest to the post-test. For men, of the five
communities identified pre-instruction, only two were identified
post-instruction. The differences resulted from the ``Force
Graph'' and ``Force Sled'' communities merging post-instruction,
possibly indicating that men developed more facility with working
with the same type of problem in multiple representations with
instruction. Pre-instruction, the ``Acceleration Graphs'' and
``Coin Toss - Acceleration'' item groups were combined; these were
separate post-instruction. These groups require different physical
principles for their solution; however, both apply the same
misconception. This may possibly indicate that men differentiate
the ideas of force and acceleration in an inconsistent manner
pre-instruction.

These results also help to explain the unfairness that was
identified in items 27-29 by Henderson \textit{et al.}
\cite{henderson2018}. Women consistently integrated this item
group (``Coin Toss - Acceleration'') with the other item groups
measuring motion under gravity (``Cart on a Ramp'' and ``Coin Toss
- Force''); men did not. ``Coin Toss - Force'' and ``Coin Toss -
Acceleration'' items differ only by asking about the force and
acceleration on a coin moving under the force of gravity; failing
to integrate the misconceptions about force and acceleration seems
to indicate either that the student does not understand that force
and acceleration are proportional or indicate some error in
interpreting the items.

The strength of the misconception, measured by the misconception
score in Table \ref{tab:misc}, shows how strongly students hold a
particular misconception. The misconception score was smaller than
the overall difference in FMCE score between men and women showing
there are not particular misconceptions more strongly held by men
or women. No gender difference in misconception score was larger
than a small effect.

\textit{RQ3: How do the communities change as the parameters of
the MMA algorithm are modified?}

Study 1 investigated variations in two network building
parameters: the correlation threshold $r$ and the community
fraction $C$. These parameters were adjusted to produce productive
community structure using the model of the correct solution
structure provided in Study 2 and the taxomony of misconceptions
provided by Hestenes and Jackson \cite{fcitable}. The threshold of
the minimum number of students who could select a response was not
investigated because productive structure was identified retaining
only responses selected by a least 30 students, the minimum
statistically viable threshold. The FMCE behaved differently; the
misconception structure changed dramatically as the threshold for
the minimum percentage of students selecting a response was
modified.

Retaining nodes selected by at least 5\% of the students, MMA
identified 35 incorrect response communities; many of these
communities were similar, with some differing by only a single
response. Retaining responses selected by at least 10\% of the
students, the structure of the communities was still complex
(Table \ref{tab:commat10misc}) but, in general, a single coherent
misconception could be identified for each community. Some, but
not all, of these misconceptions were described in the taxonomy
proposed by Hestenes, Wells, and Swackhamer \cite{hestenes1992,
fci-revised} and refined by Hestenes and Jackson \cite{fcitable}.

If responses selected by a minimum of 20\% of the students were
retained, the community structure simplified substantially (Table
\ref{tab:commat}). Examination of the community structure showed
that much of the remaining complexity involved the sporadic
inclusion of items identified as problematic by  Thornton
\textit{et al.} \cite{thornton2009comparing}. Removal of these
items produced the relatively simple community structure in Table
\ref{tab:commat2}. With the exception of one male pretest
community, these communities all measured a misconceptions
described in Hestenes and Jackson's taxonomy \cite{fcitable} as
well as requiring the same physical reasoning described in Study
2. The male pretest community applied the same misconception, but
required different physical reasoning for its correct solution.

The FCI and the FMCE community structures were dramatically
different if responses selected by 5\% of the students were
retained. At this threshold, the FCI had only 13 small communities
and the FMCE 35 often fairly large communities even though the
coverage of the FCI is substantially more broad than the FMCE.
These differences likely resulted from two sources: students in
the FCI sample scored substantially higher on the instrument than
the students in the FMCE sample and the unusual distractor
structure of the FMCE. The FCI uses only 5 responses for each
question and the incorrect responses were developed from student
interviews and include common student incorrect views. The FMCE
uses items with more than 5 responses that often generally exhaust
the possible responses. This offers far greater latitude for
students to express uncommon misconceptions and, therefore, are
only selected by a small fraction of the students.

The broad set of misconception communities identified retaining
nodes selected by 10\% of the students suggest that the state of
student incorrect reasoning may be substantially more complex than
the structure measured by the FCI.

\section{Implications}
The responses to the FCI
were constructed to measure common misconceptions allowing Jackson
and Hestenes to provide a detailed taxonomy of the misconceptions
measured by each item \cite{fcitable}. While common misconceptions
were certainly considered in the construction of the instrument,
the FMCE presents students with many possible incorrect answers.
These answers largely exhaust the possible responses. As such, the
FMCE may be a much better instrument for a purely exploratory
analysis of student incorrect thinking less tied to the
misconception view.

The identification of incorrect answer communities testing the
same misconception allows the calculation of a misconception score
as a quantitative measure of how strongly the misconception is
held. This should allow instructors to determine which
misconceptions are most prevalent in their classes and to target
instruction to eliminate these misconceptions.

\section{Limitations}
The MAMCR and MMA algorithms require a number of choices to be
made by the researcher to produce network structure that is
productive in furthering the understanding of a conceptual
instrument. As the use of network analysis matures in PER,
quantitative criteria for optimally selecting network parameters
should be developed.

\section{Conclusion}

Physics conceptual inventories have played an important role in
quantitative physics education research and understanding
students' difficulties with conceptual physics continues to be a
central research area within PER. Network analysis, specifically
Modified Module Analysis (MMA), has recently been used as a tool
to investigate the common misconceptions on the FCI
\cite{wells2019}. The current study replicated this work for the
FMCE.

In general, retaining responses selected by 20\% of the students,
the community structure for the FMCE was consistent with the item
groups identified in previous studies
\cite{thornton1998,smith2008}. The misconceptions represented by
these communities were limited: motion implies active forces,
velocity-acceleration undiscriminated, greater mass implies
greater force, and most active agent produces greatest force.
Three of these incorrect answer communities were previously
identified in the FCI \cite{wells2019}; however, the
velocity-acceleration undiscriminated misconception was only
detected as an incorrect answer community in the FMCE. The FCI was
found to measure nine misconceptions in the previous study.

The FCI and the FMCE behaved dramatically differently as network
parameters were adjusted. For the FCI, including responses
selected by 4\% of the students, only 13 communities were
detected, most with only two responses. Retaining responses
selected by a similar percentage of students, 35 communities were
detected in the FMCE with up to 15 members.

The evolution of the communities identified was dramatically
different for men and women. The communities identified for women
did not change from pretest to post-test, while only 2 of the 5
communities identified for men remained consistent. Unlike the
FCI, there was little consistency in the communities identified
for men and women either pre-instruction or post-instruction.

Overall, Modified Module Analysis was productive in understanding
the misconception structure of both the FCI and the FMCE and
allowing the comparison of the instruments.

\begin{acknowledgments}
Data collection for this work was supported by National Science
Foundation grants EPS-1003907 and ECR-1561517.
\end{acknowledgments}


%

\end{document}